\documentclass[fp,dvipdfmx,twocolumn]{jpsj3}
\usepackage[dvipdfmx]{graphicx}
\def\Sec{Sect. }
\def\sa{\alpha s }
\def\ssa{\alpha s^2 }

\def\Jeff{J_{\rm eff}}
\def\msp{m_{\rm sp}}
\def\Msp{M_{\rm sp}}

\def\v#1{\mib #1}
\def\lambdac{{\lambda_{\rm c}}}
\def\lambdaca{{\lambda_{\rm c1}}}
\def\lambdacb{{\lambda_{\rm c2}}}

\def\stot{{S^{\rm tot}}}
\def\stoto{{S_0^{\rm tot}}}

\def\H{{\mathcal H}}
\newcommand{\bra}[1]{\left\langle {#1} \right\vert}

\newcommand{\kket}[1]{\left\vert\left\vert {#1} \right.\right\rangle}
\newcommand{\ket}[1]{\left\vert {#1} \right\rangle}

\title
{
Ground-State Phase Diagram of (1/2,1/2,1) Mixed Diamond Chains
}

\author
{
Kazuo Hida\thanks{E-mail address: hida@mail.saitama-u.ac.jp}
}

\inst
{Professor Emeritus, Division of Material Science, Graduate School of Science and Engineering, \\
Saitama University, Saitama, Saitama, 338-8570\\
}

\recdate
{}

\abst
{
The ground-state phases of mixed diamond chains with ($S, \tau^{(1)}, \tau^{(2)})=(1/2,1/2,1)$, where $S$ is the magnitude of vertex spins, and $\tau^{(1)}$ and $\tau^{(2)}$ are those of apical spins, are investigated.
The two apical spins in each unit cell are coupled by an exchange coupling $\lambda$. The vertex spins are coupled with the top and bottom apical spins by exchange couplings $1+\delta$ and $1-\delta$, respectively. Although this model has an infinite number of local conservation laws for $\delta=0$, they are lost for finite $\delta$. The ground-state phase diagram is determined using the numerical exact diagonalization and DMRG method in addition to the analytical approximations in various limiting cases. The phase diagram consists of a nonmagnetic phase and several kinds of ferrimagnetic phases. We find two different ferrimagnetic phases without spontaneous translational symmetry breakdown. It is also found that the quantized ferrimagnetic phases with large spatial periodicities present for $\delta=0$ are easily destroyed by small $\delta$ and replaced by a partial ferrimagnetic phase. The nonmagnetic phase is considered to be a gapless Tomonaga-Luttinger liquid phase based on the recently extended Lieb-Schultz-Mattis theorem to the site-reflection invariant spin chains and numerical diagonalization results.}
\begin{document}
\maketitle
\section{Introduction}

In low-dimensional frustrated quantum magnets, the interplay of quantum fluctuation and frustration leads to the emergence of various exotic quantum phases.\cite{intfrust,diep} 
The conventional diamond chain\cite{Takano-K-S,ht2017} is known as one of the simplest examples in which an interplay of quantum fluctuation and frustration leads to a wide variety of ground-state phases. Remarkably, this model has an infinite number of local conservation laws, and the ground states can be classified by the corresponding quantum numbers. If the two apical spins have equal magnitudes, the pair of apical spins in each unit cell can form a nonmagnetic singlet dimer and the ground state is a direct product of the cluster ground states separated by singlet dimers.\cite{Takano-K-S,ht2017} Nevertheless, in addition to the spin cluster ground states, various ferrimagnetic states and strongly correlated nonmagnetic states such as the Haldane state are also found when the apical spins form magnetic dimers. In these cases, all the spins collectively form a correlated ground state over the whole chain. 

In the presence of various types of distortion, the spin cluster ground states also turn into highly correlated ground states. Extensive experimental studies have been also carried out on the magnetic properties of the natural mineral azurite which is regarded as an example of distorted spin-1/2 diamond chains.\cite{kiku2,kiku3}

On the other hand, the cases with unequal apical spins are less studied. In this case, the apical spins cannot form a singlet dimer. Hence, all spins in the chain inevitably form a many-body correlated state.
As a simple example of such cases, we investigated the mixed diamond chain with apical spins of magnitude 1 and 1/2, and vertex spins, 1/2 in Ref. \citen{hida2021} assuming that the exchange interactions between the vertex spins and two apical spins are equal to each other.  In the absence of coupling $\lambda$ between two apical spins, we found a quantized ferrimagnetic (QF) phase with the spontaneous magnetization per unit cell $\msp$ quantized to unity as expected from the Lieb-Mattis (LM) theorem.\cite{Lieb-Mattis} With the increase of $\lambda$, we found an infinite series of QF phases with  $\msp=1/p$, where $p$ is a positive integer ($1\leq p < \infty$) that increases with $\lambda$. Finally, the nonmagnetic Tomonaga-Luttinger liquid (TLL) phase sets in at a critical value of $\lambda=\lambdac$. The width and spontaneous magnetization of each QF phase tend to infinitesimal as $\lambda$ approaches $\lambdac$. 

If the two apical spins have different magnitudes, however, it is natural to assume that the exchange interactions between these two kinds of apical spins and the vertex spins are also different. We examine this case in the present work. 
 Two QF phases without translational symmetry breakdown are found. The QF phases with large $p$ are replaced by a partial ferrimagnetic (PF) phase in which the magnetization varies continuously with the exchange parameter. With the help of numerical diagonalization results, the nonmagnetic phase is considered to be a TLL phase consistent with the Lieb-Schultz-Mattis (LSM) theorem\cite{LSM,Tasaki2018,fuji2016, Po2017,OTT2021} that is recently extended to site-reflection invariant spin chains\cite{fuji2016, Po2017,OTT2021}.

This paper is organized as follows.
{In \Sec 2}, the model Hamiltonian is presented.
{In \Sec 3}, various limiting cases are examined analytically. In \Sec 4, the classical ground state is analytically determined. In \Sec 5, the ground-state phase diagram determined by the numerical calculation is presented and the properties of each phase are discussed.
The last section is devoted to a summary and discussion.

\section{Model}

\begin{figure}[th]
\centerline{\includegraphics[width=7cm]{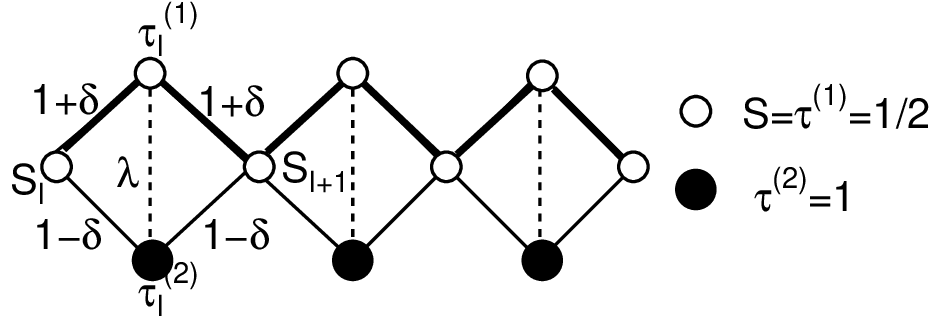}}
\caption{Structure of the diamond chain investigated in this work.}
\label{lattice}
\end{figure}

We investigate the ground-state phases of mixed diamond chains described by the following Hamiltonian:
\begin{align}
{\mathcal H} &= \sum_{l=1}^{L} \Big[(1+\delta)\v{S}_{l}(\v{\tau}^{(1)}_{l}+\v{\tau}^{(1)}_{l-1})
\nonumber\\
&+(1-\delta)\v{S}_{l}(\v{\tau}^{(2)}_{l}+\v{\tau}^{(2)}_{l-1})
+ \lambda\v{\tau}^{(1)}_{l}\v{\tau}^{(2)}_{l}\Big],
\label{hama}
\end{align}
where $\v{S}_{l}, \v{\tau}^{(1)}_{l}$ and $\v{\tau}^{(2)}_{l}$ are spin operators with magnitudes ${S}_{l}={\tau}^{(1)}_{l}=1/2$ and ${\tau}^{(2)}_{l}=1$, respectively. The number of unit cells is denoted by $L$, and the total number of sites is $3L$. 
The lattice structure is depicted in Fig. \ref{lattice}.
We consider the region $\lambda \geq 0$ and $1 \geq \delta\geq -1$. For $\delta=0$, $(\v{\tau}^{(1)}_l+\v{\tau}^{(2)}_l)^2$ commutes with the Hamiltonian (\ref{hama}) for all $l$. In Ref. \citen{hida2021}, we made use of this property to determine the ground-state phase diagram.
In the present work, we examine the general case of $\delta \neq 0$.

\section{Analytical Results}

\begin{figure}[th]
\centerline{\includegraphics[width=6cm]{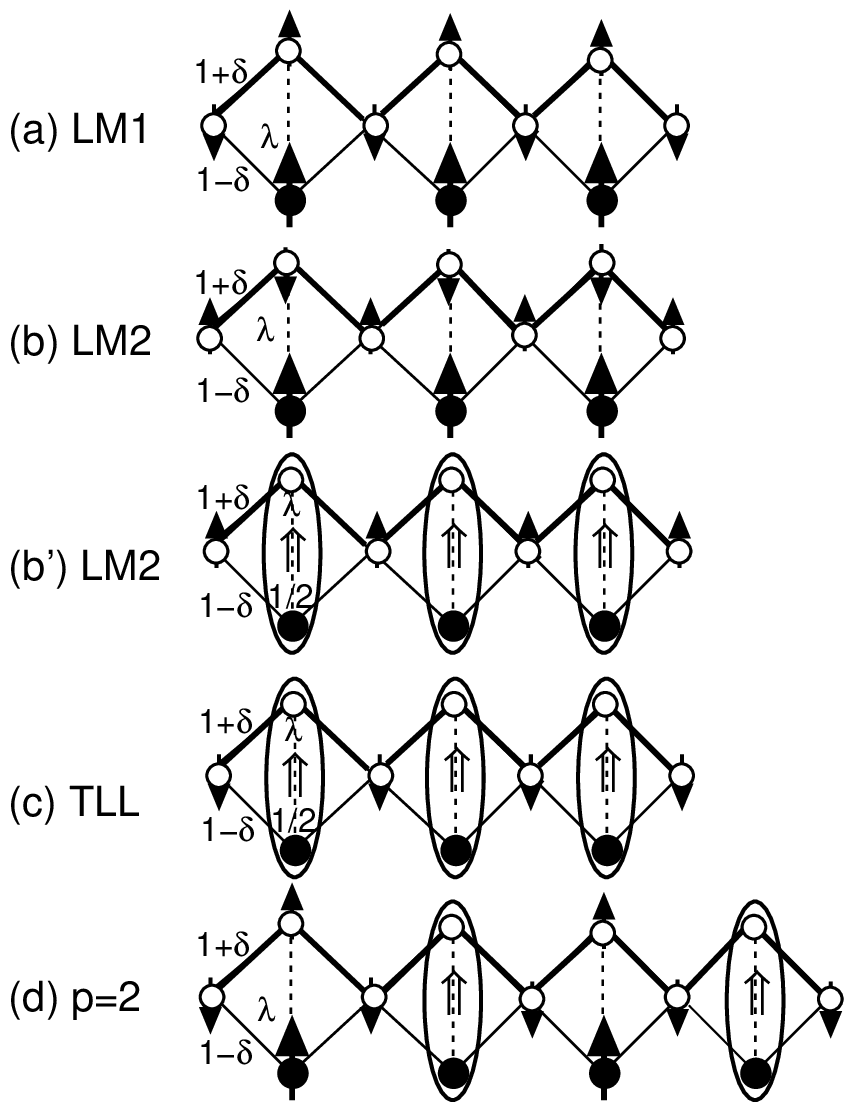}}
\caption{Schematic spin configurations in the ground-state phases. The ovals in (b'), (c), and (d) mean that the two spins in an oval form a spin-doublet state. The open arrows express the total spins $\v{T}_l$ in ovals. In (c), the antiferromagnetic long-range order depicted in the figure melts due to the quantum fluctuation resulting in the nonmagnetic TLL.}
\label{fig:conf}
\end{figure}
\subsection{Lieb-Mattis theorem}
We start with several limiting cases where we can examine the ground state analytically. 
Before going into the discussion of specific cases, we briefly introduce the Lieb-Mattis theorem which is useful to determine the spontaneous magnetization in the absence of frustration.
 Let us consider a Heisenberg model
\begin{align}
\H_{\rm H}=\sum_{i,j}J_{ij}\v{S}_i\v{S}_j
\end{align}
defined on a lattice consisting of two sublattices A and B. Here, $\v{S}_i$ is the spin operator on the $i$-th site with arbitrary magnitude. It is assumed that the magnetic interactions $J_{ij}$ satisfy the following condition:
\begin{enumerate}
\item If two spins are on different sublattices, then $J_{ij} \geq 0$.
\item If two spins are on the same sublattice, then $J_{ij} \leq 0$.
\item All spins are connected by magnetic interaction. 
\end{enumerate}
These assumptions imply the absence of frustration. Then, the spontaneous magnetization $\Msp$ of the ground state is given by $\Msp=| S_{\rm A}-S_{\rm B}|$, where $S_{\rm A}$ and $S_{\rm B}$ are the sums of the magnitudes of the spins on the A-sublattice  and B-sublattice, respectively.\cite{Lieb-Mattis}

\subsection{$\lambda=0$}

If $-1 < \delta < 1 $, the system is unfrustrated and the ground state is the QF phase with $\msp=1$ according to the LM theorem.\cite{Lieb-Mattis} In this case, the sublattice A consists of the sites occupied by $\v{S}_l$, and the sublattice B, those occupied by $\v{\tau_l^{(1)}}$ and  $\v{\tau_l^{(2)}}$. Hence, $S_{\rm A}=L/2$ and $S_{\rm B}=3L/2$ which gives $\Msp=L$ and $\msp=\Msp/L=1$. The numerical analysis in \Sec \ref{sec:num} shows that this phase survives even in the weakly frustrated regime of small $\lambda$. Hereafter, this phase is called the LM1 phase.  Schematic spin configuration is presented in Fig. \ref{fig:conf}(a).

\subsection{$\delta=1$}

If $\lambda > 0$, the ground state is the QF phase with $\msp=1$ according to the LM theorem.  In this case, the sublattice A consists of the sites occupied by $\v{\tau_l^{(1)}}$, and the sublattice B, those occupied by $\v{S}_l$ and  $\v{\tau_l^{(2)}}$. Hence, $S_{\rm A}=L/2$ and $S_{\rm B}=3L/2$ which gives $\Msp=L$ and $\msp=\Msp/L=1$. The numerical analysis in \Sec 5 shows that this phase survives even in the weakly frustrated regime of small $1-\delta$. Hereafter, this phase is called the LM2 phase. The schematic spin configuration presented in Fig. \ref{fig:conf}(b) demonstrates that this phase is distinct from the LM1 phase. This is also numerically confirmed in \Sec \ref{sec:num}.

\subsection{$\lambda \gg 1+\delta, 1-\delta$}\label{item:larlam}

In the limit $\lambda \rightarrow \infty$, all pairs of $\v{\tau}^{(1)}_l$ and $\v{\tau}^{(2)}_l$ form doublet states $\kket{T_l,T_l^z}$ with $T_l=1/2$ and $T_l^z=\pm 1/2$ where $\v{T}_l\equiv \v{\tau}^{(1)}_l+\v{\tau}^{(2)}_l$. They are expressed using the basis $\ket{{\tau}^{(1)z}_l, {\tau}^{(2)z}_l}$ as
\begin{align}
\kket{\dfrac{1}{2},\dfrac{1}{2}}&=\sqrt{\frac{1}{3}}\ket{0,\dfrac{1}{2}}-\sqrt{\frac{2}{3}}\ket{1,-\dfrac{1}{2}},\label{eq:doub1}\\
\kket{\dfrac{1}{2},-\dfrac{1}{2}}&=\sqrt{\frac{1}{3}}\ket{0,-\dfrac{1}{2}}-\sqrt{\frac{2}{3}}\ket{-1,\dfrac{1}{2}}.
\end{align}
For large but finite $\lambda$,  $\v{T}_l$ and $\v{S}_{l'}$ ($l'=l$ or $l+1$) are coupled by an effective Heisenberg interaction $J_{\rm eff}$. If $J_{\rm eff}$ is antiferromagnetic, the ground state is the nonmagnetic TLL state as schematically shown in Fig. \ref{fig:conf}(c). If it is ferromagnetic, the ground state is the QF state with $\msp=1$ as shown in Fig. \ref{fig:conf}(b'). This ground state configuration is continuously deformed to that of the LM2 phase depicted in Fig. \ref{fig:conf}(b) by reducing the coefficient of $\ket{0,1/2}$ in Eq. (\ref{eq:doub1}) continuously. Hence, this QF state also belongs to the LM2 phase. We estimate $\Jeff$ within the first order perturbation calculation with respect to $(1\pm\delta)/\lambda$ as
\begin{align}
\Jeff=\frac{1}{3}(3-5\delta).
\end{align}
Hence, we find the phase boundary between the TLL phase and the LM2 phase is given by $\delta=0.6$ for large enough $\lambda$.

\subsection{$\delta=-1$}

The system is unfrustrated and the ground state is the nonmagnetic phase with $\msp=0$ according to the LM theorem.  In this case, the sublattice A consists of the sites occupied by $\v{\tau_l^{(2)}}$, and the sublattice B, those occupied by $\v{S}_l$ and  $\v{\tau_l^{(1)}}$. Hence, $S_{\rm A}=L$ and $S_{\rm B}=L$ which gives $\Msp=0$ and $\msp=\Msp/L=0$.  According to the numerical calculation of \Sec \ref{sec:num}, this phase continues to the TLL phase discussed in \Sec \ref{item:larlam}.

\subsection{$1+\delta\simeq 0$ and $\lambda\simeq 0$}

For $1+\delta=\lambda=0$, $\v{S_l}$ and $\v{\tau}_l^{(2)}$ form a ferrimagnetic chain with $\msp=1/2$, and $\v{\tau}^{(1)}_l$ are free spins with magnitude $1/2$.
We take the zeroth order Hamiltonian as
\begin{align}
{\mathcal H}_0 &= \sum_{l=1}^{L} \Big[(1-\delta)\v{S}_{l}(\v{\tau}^{(2)}_{l}+\v{\tau}^{(2)}_{l-1})
\Big],\label{ham0}
\end{align}
and the total spin of ${\mathcal H}_0$ is defined by
\begin{align}
\v{S}^{\rm tot}_0=\sum_{l=1}^{L} (\v{S}_l+\v{\tau}^{(2)}_{l}).
\end{align}
The perturbation part of the Hamiltonian is rewritten as,
\begin{align}
{\mathcal H}_1&= \sum_{l=1}^{L} \Big[(1+\delta)(\v{S}_{l}+\v{S}_{l+1})+ \lambda\v{\tau}^{(2)}_{l}\Big]\v{\tau}^{(1)}_{l}.\label{ham1}
\end{align}
Within the subspace of the ferrimagnetic ground states of ${\mathcal H}_0$ with $\stoto=L/2$, each ground state is specified by ${\stoto}^z=-L/2,..., L/2$. From rotational symmetry, the two-spin effective Hamiltonian ${\mathcal H}_l^{\rm eff}$ for $\v{S}^{\rm tot}_0$ and $\v{\tau}^{(1)}_l$ is written down by their inner product as
\begin{align}
{\mathcal H}_l^{\rm eff} = & \frac{\Jeff '}{(L/2)}\v{S}_0^{\rm tot}\v{\tau}^{(1)}_{l}.
\end{align}
It should be noted that the terms of higher powers of $\v{S}_0^{\rm tot}\v{\tau}^{(1)}_{l}$ are unnecessary since  $\v{S}_0^{\rm tot}\v{\tau}^{(1)}_{l}$ can take only two values  ${S}_0^{\rm tot}/2$ and $-({S}_0^{\rm tot}+1)/2$.  Then, the total effective Hamiltonian ${\mathcal H}^{\rm eff}$ is given by
\begin{align}
{\mathcal H}^{\rm eff} = & \sum_{l=1}^{L}{\mathcal H}_l^{\rm eff}.
\end{align}
To determine $\Jeff '$, we estimate the expectation values of ${\mathcal H}_1$ and ${\mathcal H}^{\rm eff}$ in the state $\ket{F}=\ket{S^{\rm tot}_0=L/2; S^{{\rm tot} z}_0=L/2}$ to find
\begin{align}
\bra{F}{\mathcal H}^{\rm eff}\ket{F} = &\sum_{l=1}^{L} \frac{2\Jeff '}{L}\bra{F}{S}_0^{{\rm tot}z}\ket{F}{\tau}^{(1)z}_{l}\nonumber\\
&= \Jeff '\sum_{l=1}^{L} {\tau}^{(1)z}_{l}, \label{eq:heff_exp}\\
\bra{F}{\mathcal H}_1\ket{F}= &\sum_{l=1}^{L} \Big[2(1+\delta)\bra{F}{S}^z_{l}\ket{F}\nonumber\\
&+ \lambda\bra{F}{\tau}^{(2)z}_{l}\ket{F}\Big]{\tau}^{(1)z}_{l}.\label{eq:h1_exp}
\end{align}
Comparing (\ref{eq:heff_exp}) and (\ref{eq:h1_exp}),  we find
\begin{align}
\Jeff ' &=2(1+\delta)\bra{F}{S}^z_{l}\ket{F}+ \lambda\bra{F}{\tau}^{(2)z}_{l}\ket{F}.
\end{align}

The expectation values $\bra{F}{S}^z_{l}\ket{F}$ and $\bra{F}{\tau}^{(2)z}_{l}\ket{F}$ are calculated by the numerical diagonalization for $L=2,4,6,12$ and 14. After two steps of Shanks transformation,\cite{shanks} we find
\begin{align}
\bra{F}{S}^z_{l}\ket{F} &\simeq 0.7924871,\\ \bra{F}{\tau}^{(2)z}_{l}\ket{F}&\simeq -0.2924871.
\end{align}

Thus, we can determine the phase boundary between the LM1 phase and the nonmagnetic phase as $\lambda/(1+\delta)\simeq 0.73815$ where the sign of $\Jeff '$ changes.

\section{Classical Phase Diagram}

\begin{figure}[th]
\centerline{\includegraphics[width=4cm]{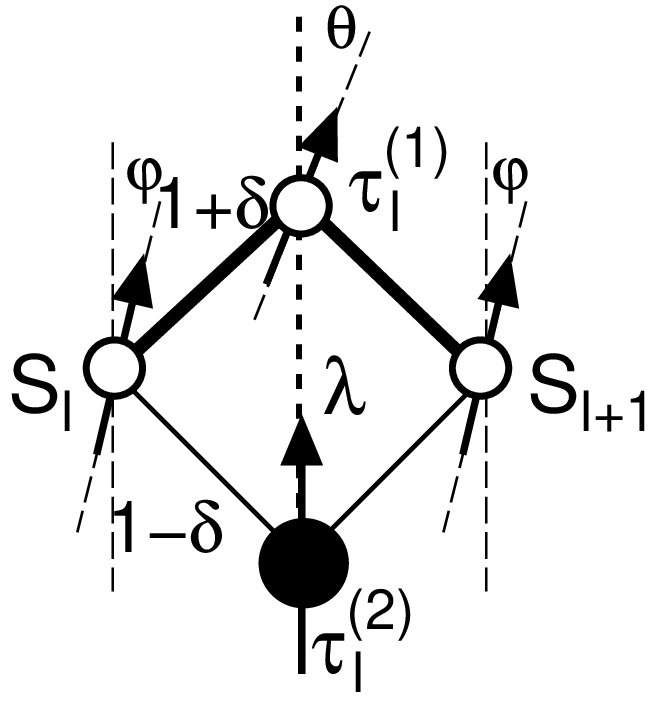}}
\caption{Definition (\ref{eq:angles}) of spin angles $\theta$ and $\varphi$.}
\label{angles}
\end{figure}

Before the description of the numerical phase diagram, we examine the classical limit. We regard all spins as classical vectors with fixed magnitudes. The magnitudes of $\v{S}_l$ and $\v{\tau}^{(1)}_l$ are denoted by $s$, and that of $\v{\tau}^{(2)}_l$, by $\alpha s$. We assume a uniform ground-state spin configuration in the form,
\begin{align}
\begin{split}
\v{S}_{l}&=(0,s\sin\varphi,s\cos\varphi),\\
\v{\tau}^{(1)}_{l}&=(0,s\sin\theta,s\cos\theta),\\
\v{\tau}^{(2)}_{l}&=(0,0,\sa),\label{eq:angles}
\end{split}
\end{align}
as depicted in Fig. \ref{angles}. We take the direction of $\v{\tau}^{(2)}_{l}$ as $z$-direction.
The nonuniform configurations such as period-doubled states and spiral states are also considered. However, they turned out to have higher energies.

\begin{figure}[th]
\centerline{\includegraphics[width=6cm]{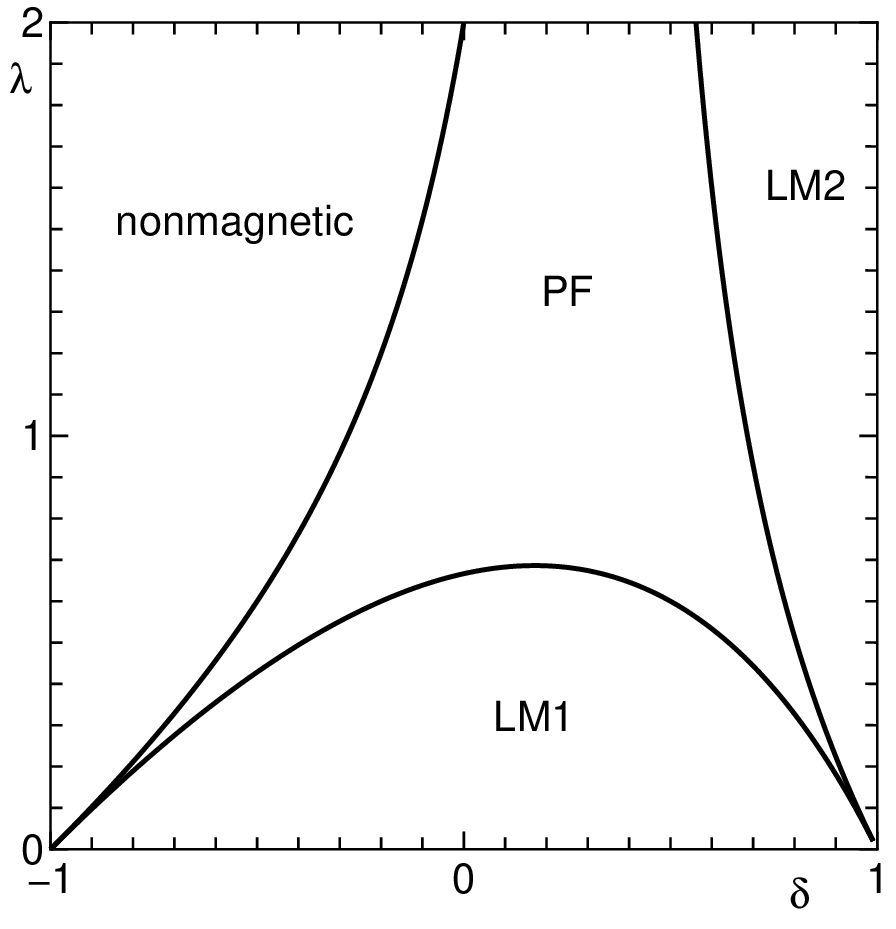}}
\caption{Ground-state phase diagram in the classical limit with $\sa=1$ and $s=1/2$.}
\label{fig:class}
\end{figure}

The ground-state energy per unit cell is given by
\begin{align}
E&=2s^2(1+\delta)\cos(\theta-\varphi)\nonumber\\
&+2\ssa(1-\delta)\cos\varphi+\ssa\lambda\cos\theta.
\end{align}
Minimizing $E$ with respect to $\theta$ and $\varphi$, we have
\begin{align}
\frac{\partial E}{\partial \theta}&=-2s^2(1+\delta)\sin(\theta-\varphi)-\ssa\lambda\sin\theta=0,\label{eq:et}\\
\frac{\partial E}{\partial \varphi}&=2s^2(1+\delta)\sin(\theta-\varphi)\nonumber\\
&-2\ssa(1-\delta)\sin\varphi=0.\label{eq:ep}
\end{align}
Let us start with trivial solutions.
\begin{enumerate}
\item $\theta=\varphi=0$
\begin{align}
\begin{split}
\v{S}_{l}&=(0,0,s),\\
\v{\tau}^{(1)}_{l}&=(0,0,s),\\
\v{\tau}^{(2)}_{l}&=(0,0,\sa).
\end{split}
\end{align}
This is a ferromagnetic phase with spontaneous magnetization $\msp=(2+\alpha)s$ per unit cell. This phase is not realized in the parameter regime considered ($\lambda >0$ and $-1 \leq \delta \leq 1$).
\item $\theta=\varphi=\pi$ :
\begin{align}
\begin{split}
\v{S}_{l}&=(0,0,-s),\\
\v{\tau}^{(1)}_{l}&=(0,0,-s),\\
\v{\tau}^{(2)}_{l}&=(0,0,\sa).
\end{split}
\end{align}
For $\alpha=2$, this is a N\'eel-type ground state with long-range antiferromagnetic order. However, once the quantum fluctuation is switched on, it is expected that this state turns into the nonmagnetic phase with $\msp=0$ owing to the one-dimensionality. Hence, this corresponds to the classical counterpart of the nonmagnetic phase.
\item $\theta=0, \varphi=\pi$ :
\begin{align}
\begin{split}
\v{S}_{l}&=(0,0,-s),\\
\v{\tau}^{(1)}_{l}&=(0,0,s),\\
\v{\tau}^{(2)}_{l}&=(0,0,\sa).
\end{split}
\end{align}
This is the classical counterpart of the LM1 phase with spontaneous magnetization $\msp=\sa$ per unit cell.
\item $\theta=\pi, \varphi=0$ :
\begin{align}
\begin{split}
\v{S}_{l}&=(0,0,s),\\
\v{\tau}^{(1)}_{l}&=(0,0,-s),\\
\v{\tau}^{(2)}_{l}&=(0,0,\sa).
\end{split}
\end{align}
This is the classical counterpart of the LM2 phase with spontaneous magnetization $\msp=\sa$ per unit cell.
\item Nontrivial solution :

Assuming $\sin\theta\neq 0$ and $\sin\varphi\neq 0$, we find
\begin{align}
\cos\varphi&=\frac{\lambda(1+\delta)}{4\alpha(1-\delta)^2}-\frac{\alpha\lambda}{4(1+\delta)}-\frac{(1+\delta)}{\alpha\lambda},\label{eq:cosp}\\
\cos\theta&=\frac{2(1-\delta^2)}{\alpha\lambda^2}-\frac{(1+\delta)}{2\alpha(1-\delta)}-\frac{\alpha(1-\delta)}{2(1+\delta)},\label{eq:cost}
\end{align}
after some elementary manipulations from (\ref{eq:et}) and (\ref{eq:ep}). The spontaneous magnetization $\msp$ per unit cell is given by
\begin{align}
&\msp^2=(\sa)^2+2\ssa(\cos\theta+\cos\varphi)\nonumber\\
&+2s^2\left(1+\cos\varphi\cos\theta-\frac{\lambda}{2(1-\delta)}(1-\cos^2\theta)\right).
\end{align}
This state corresponds to the PF phase.
\end{enumerate}
The phase boundary between the PF phase and other phases can be obtained in the following way, 
\begin{enumerate}
\item PF-LM1 ($\theta=0, \varphi=\pi$) phase boundary :

Setting $\cos\theta=1$ in (\ref{eq:cost}), we have $\lambda=\lambdaca$ where
\begin{align}
\lambdaca&=\frac{2(1-\delta^2)}{\alpha(1-\delta)+(1+\delta)}.\label{eq:lambdaca}
\end{align}

The value of $\cos\varphi$ at $\lambda=\lambdaca$ is obtained by substituting (\ref{eq:lambdaca}) into (\ref{eq:cosp}) as
\begin{align}
&\cos\varphi= -1.
\end{align}
This implies that $\lambdaca$ corresponds to the PF-LM1 phase boundary.

\item PF-LM2 ($\theta=\pi, \varphi=0$) and PF-nonmagnetic ($\theta=\varphi=\pi$) phase boundary :

Setting $\cos\theta=-1$ in (\ref{eq:cost}), we have $\lambda=\pm\lambdacb$ where
\begin{align}
\lambdacb&=\frac{2(1-\delta^2)}{(1+\delta)-\alpha(1-\delta)}.\label{eq:lambdacb}
\end{align}
The value of $\cos\varphi$ at $\lambda=\pm\lambdacb$ is obtained by subsstituting (\ref{eq:lambdacb}) into (\ref{eq:cosp}) as
\begin{align}
\cos\varphi&=\pm1.
\end{align}
This implies that $\lambdacb$ and $-\lambdacb$ correspond to the PF-LM2 and PF-nonmagnetic phase boundary, respectively.
\end{enumerate}
The classical phase diagram obtained in this section is shown in Fig. \ref{fig:class} for $\sa=1$ and $s=1/2$.
\begin{figure}[h]
\centerline{\includegraphics[width=6cm]{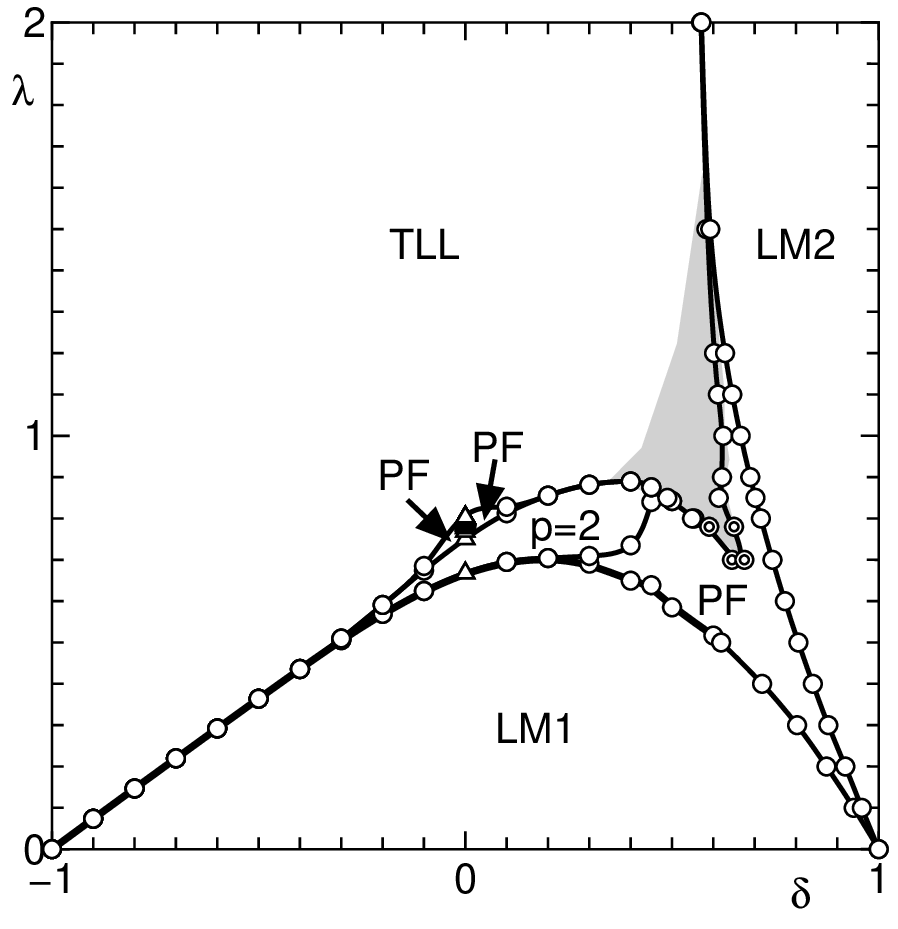}}
\caption{Ground-state phase diagram. The open circles are the phase boundaries estimated from the numerical exact diagonalization data extrapolated to the thermodynamic limit from $L=4,6$ and 8 by the Shanks transform.\cite{shanks} The double circles are the phase boundaries estimated from the DMRG data for $L=48$. The triangles at $\delta=0$ are the phase boundaries between the infinite series of QF phases determined in Ref. \citen{hida2021}. The QF phases with $p > 2$ for finite $\delta$ are not shown, since they survive only for invisibly small $\delta$ in the present scale. The deviation from the scaling relation $\Delta E \sim 1/L$ for $L =18$ and 24 is significant in the shaded area. The curves are guides for the eye.}
\label{fig:phase}
\end{figure}
\section{Numerical Results}\label{sec:num}

We have carried out the numerical exact diagonalization for $L=4,6$ and 8 with the periodic boundary condition to determine the phase boundary from the values of the spontaneous magnetization. The extrapolation of the transition point to the thermodynamic limit is carried out using the Shanks transform.\cite{shanks} If the data for larger systems are necessary, the DMRG calculation for $L=48$ is carried out with the open boundary condition. The obtained phase diagram is shown in Fig. \ref{fig:phase}.

\subsection{Ferrimagnetic phases with $\msp=1$}
\begin{figure}[th]
\centerline{\includegraphics[height=6cm]{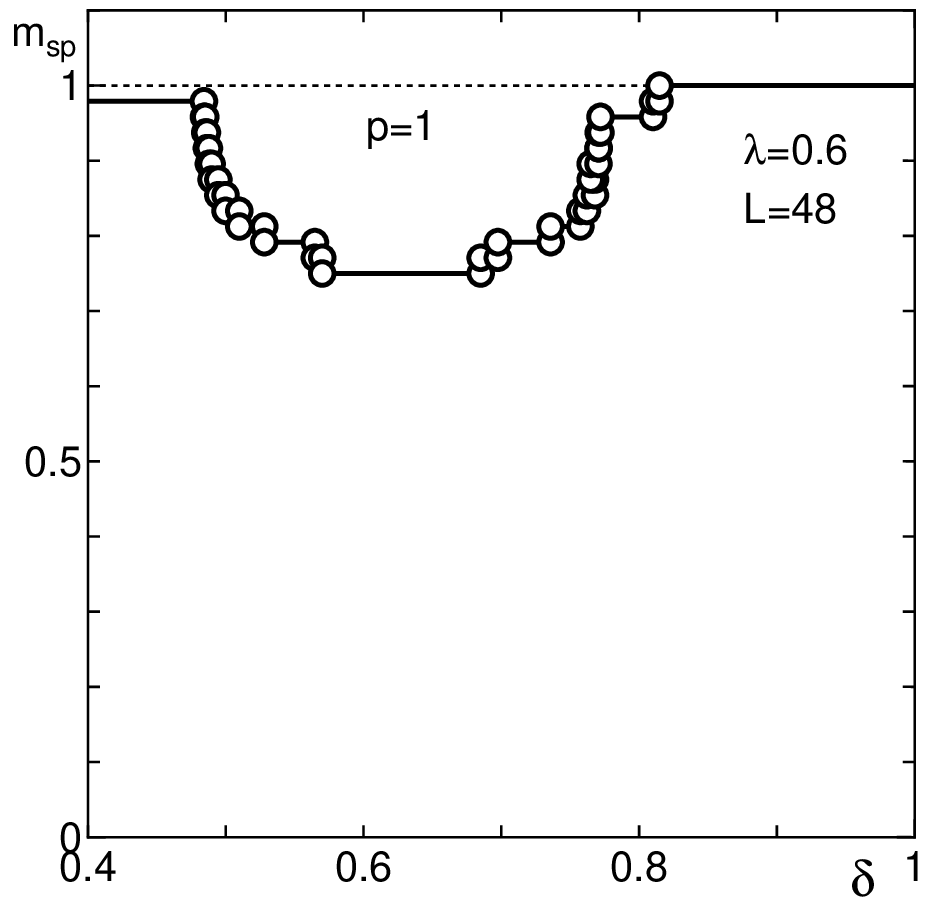}}
\caption{$\delta$-dependence of $\msp$ in the ground state for $\lambda=0.6$ calculated by the DMRG method for open chains with $L=48$. }
\label{spmag_lp06}
\end{figure}

\begin{figure}[th]
\centerline{\includegraphics[height=6cm]{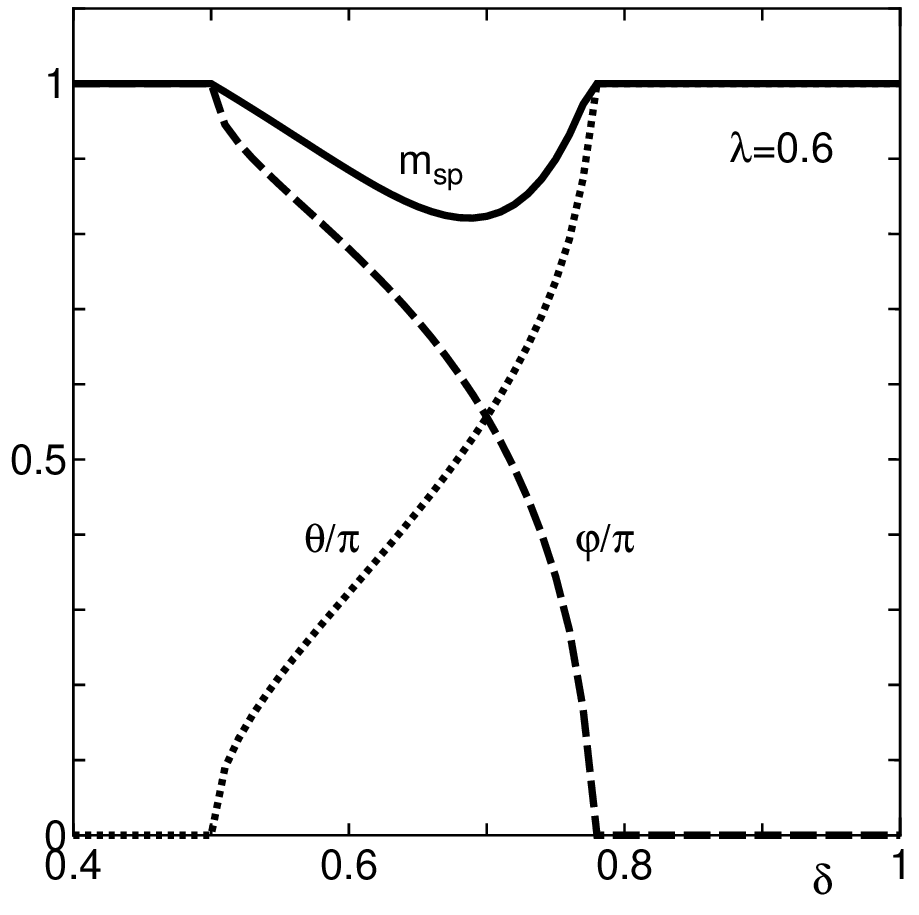}}
\caption{$\delta$-dependence of $\msp$ in the classical ground state for $\lambda=0.6$. The angles $\theta$ and $\varphi$ are also plotted. }
\label{smagclas_l06}
\end{figure}

As in the classical case, the two QF phases (LM1, LM2) with $\msp=1$ do not form a single phase but are separated by the PF phase, the TLL phase, and the $p=2$ QF phase. The $\delta$-dependence of the spontaneous magnetization $\msp$ for $L=48$ calculated by the DMRG method is shown in Fig. \ref{spmag_lp06} for $\lambda=0.6$. This behavior shows that a PF phase with $\msp < 1$ intervenes between the two QF phases with $\msp=1$. The corresponding behavior in the classical limit is also shown in Fig. \ref{smagclas_l06}. The angles $\theta$ and $\varphi$ vary by $\pi$ across the PF phase. This behavior explicitly shows that the LM1 and LM2 phases are different phases.

\begin{figure}[th]
\centerline{\includegraphics[height=6cm]{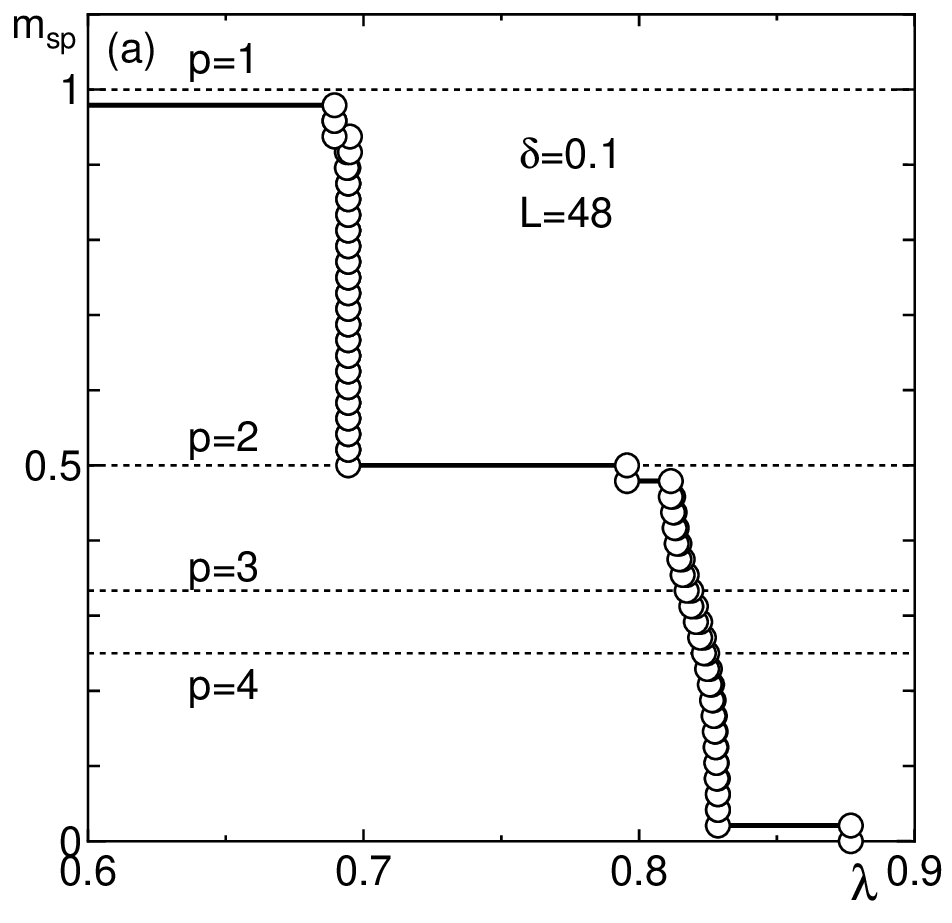}}
\centerline{\includegraphics[height=6cm]{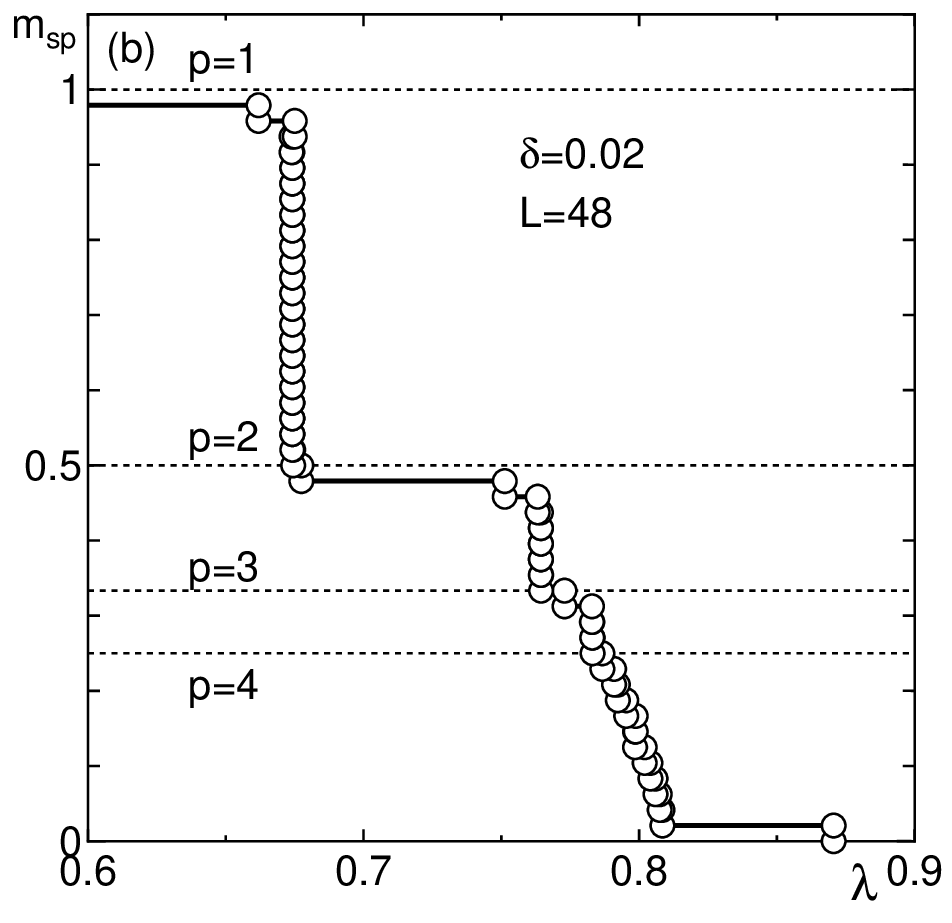}}
\centerline{\includegraphics[height=6cm]{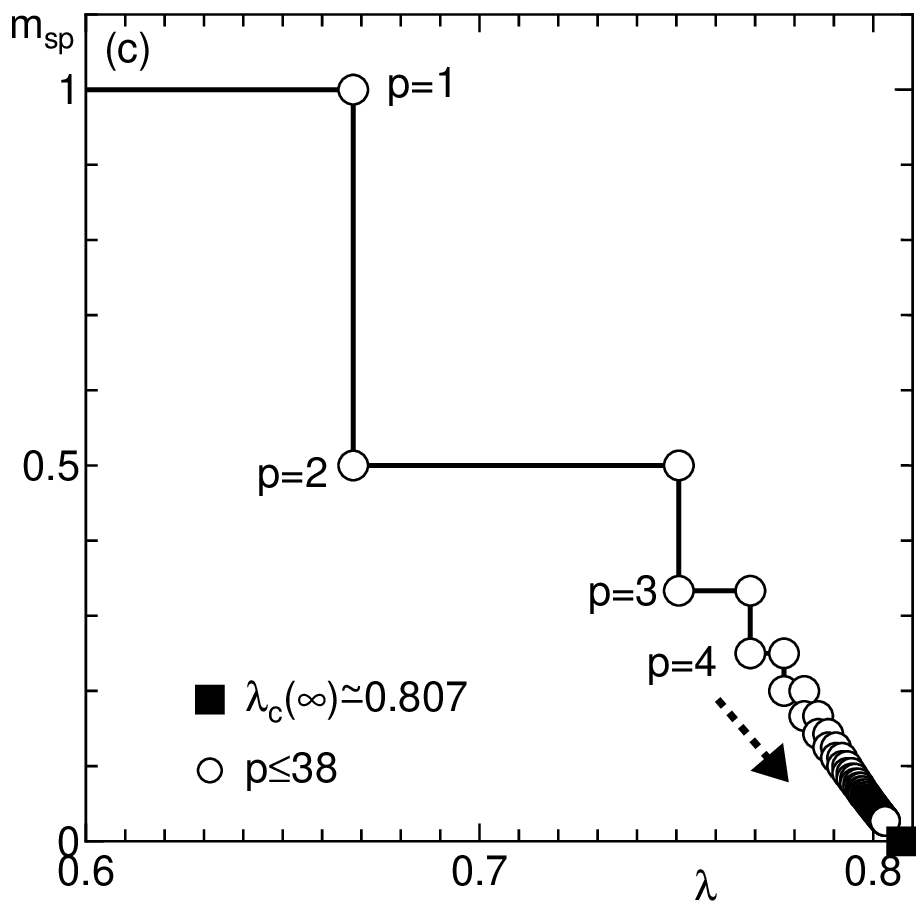}}
\caption{$\lambda$-dependence of the spontaneous magnetization in the ground state for (a) $\delta=0.1$ and (b) $\delta=0.02$ calculated by the DMRG method for open chains with $L=48$. (c) The corresponding figure for $\delta=0$ is taken from Ref. \citen{hida2021}.}
\label{spmag_p01}
\end{figure}
\begin{figure}[th]
\centerline{\includegraphics[height=6cm]{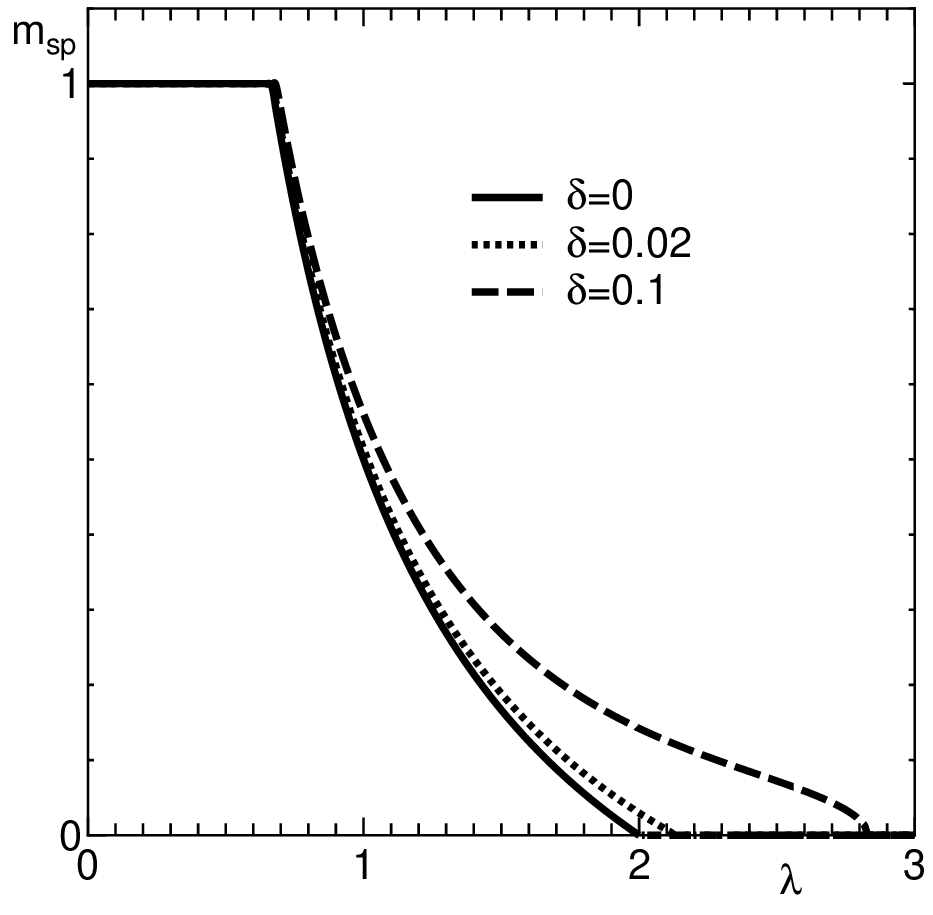}}
\caption{$\lambda$-dependence of the spontaneous magnetization in the ground state for $\delta=0$, 0.02 and 0.1 in the classical limit.}
\label{spmag_clas}
\end{figure}

\subsection{The fate of the infinite series of QF phases}

Figure \ref{spmag_p01} shows the $\lambda$-dependence of the spontaneous magnetization for (a) $\delta=0.1$ and (b) 0.02. The corresponding figure for $\delta=0$ taken from Ref. \citen{hida2021} that shows the presence of an infinite series of QF phases is also shown as Fig. \ref{spmag_p01}(c) for comparison. For $\delta=0.02$, the QF phases with $\msp=1$ $(p=1)$ and $\msp=1/2$ $(p=2)$ remain finite and the structures survive around the magnetizations corresponding to $p =3$ and 4. For $\delta=0.1$, only the QF phases with $\msp=1$ $(p=1)$ and $\msp=1/2$ $(p=2)$ remain finite, while those with $p \geq 3$ are smeared out and replaced by a PF phase. These results suggest that the QF phases become more fragile with increasing $p$. The corresponding curves in the classical limit are shown in Fig. \ref{spmag_clas}. The QF phases vanish in the classical case showing that these are essentially quantum effects.

The fragility of the QF phase with large $p$ can be understood in the following way: If the spontaneous breakdown of the translational invariance were absent, the QF ground states with $p \geq 2$ are absent and the ground state is a PF state that is regarded as a magnetized TLL.\cite{furuya_giamarchi}  The low energy effective Hamiltonian in this phase is given by the $U(1)$ compactified boson field theory with the TLL parameter $K$ as follows:
\begin{align}
\H_{\rm B}^{(0)}={1\over 2\pi}\int dx \left[K(\pi \Pi)^2+\left({1\over K}\right)
(\partial_x \phi)^2 \right],
\end{align}
where $\phi$ is a boson field defined on a circle $\phi \in [0, \sqrt{2}\pi)$, and $\Pi$ is the momentum density field conjugate to $\phi$. The spin wave velocity is set equal to unity. Extending the bosonization procedure of Ref. \citen{oya} to the case of mixed spin chains, a translation by one unit cell results in the shift of the boson field as
\begin{align}
\phi \rightarrow \phi+\sqrt{2}\pi(S_{\rm uc}- \msp),\label{eq:trans}
\end{align}
where $S_{\rm uc}$ is the sum of the spin magnitudes in a unit cell. In the present case, $S_{\rm uc}=2$. We consider the case $\msp=1/p$ that corresponds to the value of $\msp$ in the QF phase with period $p$. Then, taking the compactification condition into account, the shift (\ref{eq:trans}) is rewritten as 
\begin{align}
\phi \rightarrow \phi-\sqrt{2}\pi/p. \label{eq:shift}
\end{align}
The leading perturbation invariant under the shift (\ref{eq:shift}) is given by
\begin{align}
\H_{\rm B}^{(1)}=c \int_{-\infty}^{\infty}dx\cos(\sqrt{2}p\phi),\label{eq:perturb}
\end{align}
where $c$ is a constant.  Although this operator is translationally invariant, if it is relevant, the phase $\phi$ is pinned to one of the minima of $c\cos(\sqrt{2}p\phi)$ and the translational invariance is spontaneously broken. Since the scaling dimension of the operator (\ref{eq:perturb}) is $x_p=p^2K/2$, it is relevant if $x_p <2$. Although the $p$-dependence of $K$ is unknown, assuming that it is moderate, the main $p$ dependence of $x_p$ comes from the factor of $p^2$. This explains why the QF phases with large $p$ are more fragile than those with small $p$.

\subsection{Nonmagnetic phase}
\begin{figure}[th]
\centerline{\includegraphics[height=6cm]{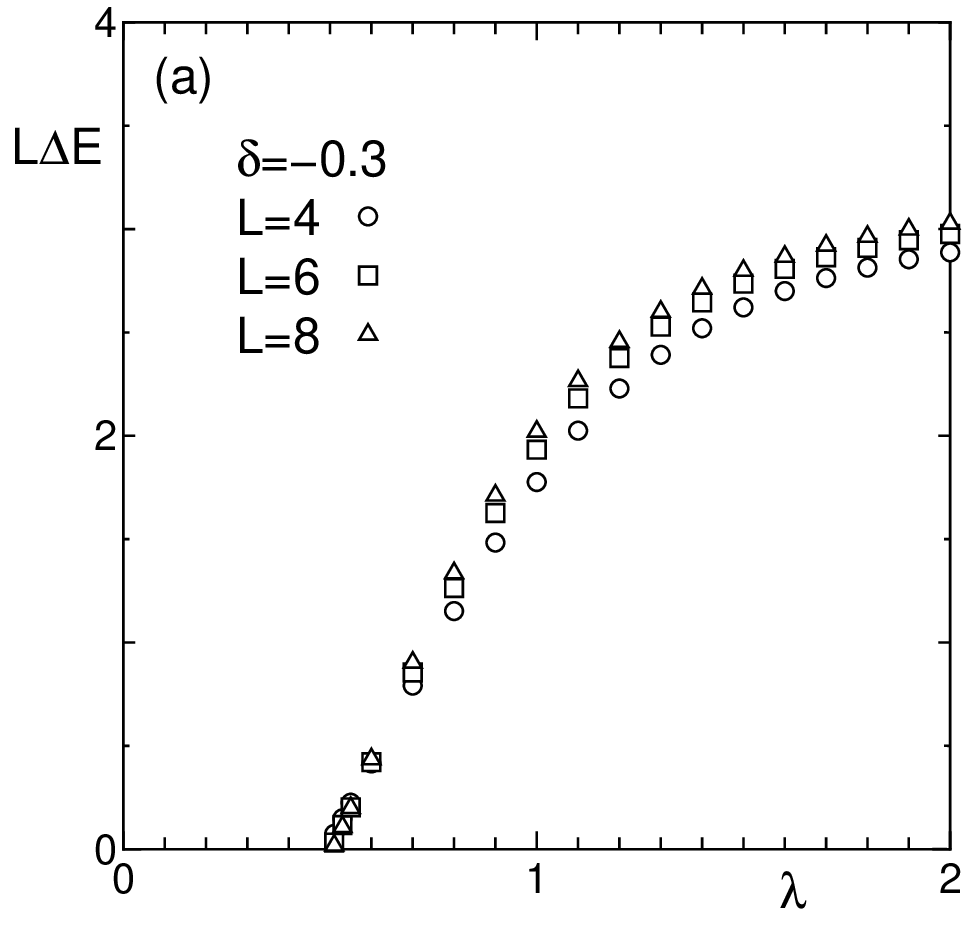}}
\centerline{\includegraphics[height=6cm]{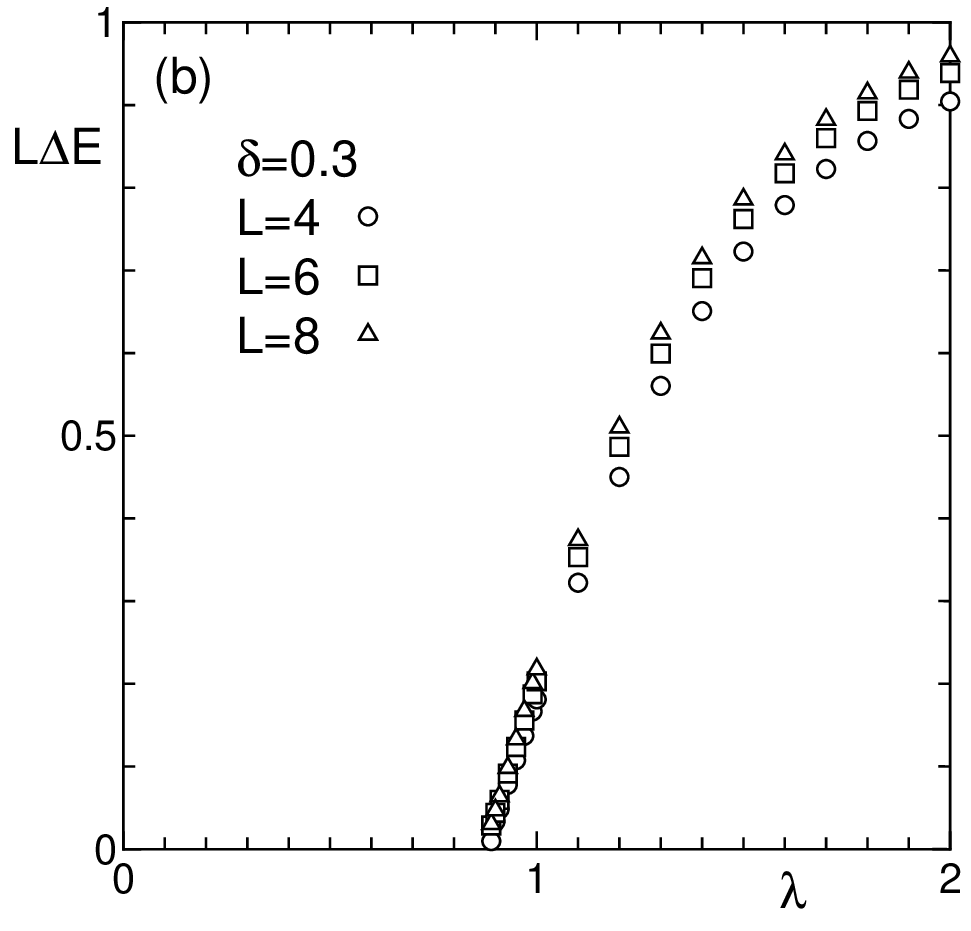}}
\centerline{\includegraphics[height=6cm]{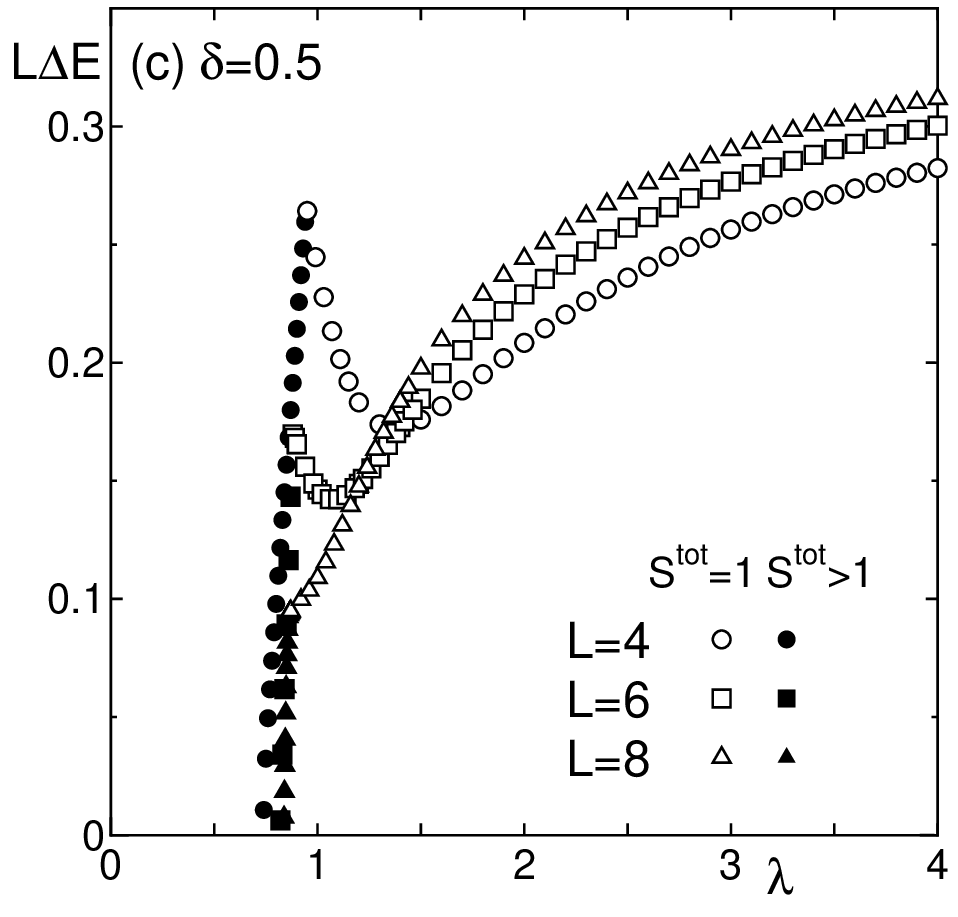}}
\caption{$\lambda$-dependence of the scaled gap $L\Delta E$ of the lowest excitations for (a) $\delta=-0.3$, (b) 0.3 and (c) 0.5 with $L=4, 6$ and 8. The open and filled symbols are excitations with $\stot=1$ and $\stot > 1$, respectively.}
\label{fig:scgap}
\end{figure}

For larger $\lambda$, the nonmagnetic phase appears and it continues to the TLL phase for large $\lambda$ discussed in \Sec \ref{item:larlam}. Since the sum of the spin magnitudes in a unit cell is an integer, the conventional LSM theorem\cite{LSM, Tasaki2018} does not exclude the unique gapped phase. However, our model (\ref{hama}) is invariant under the site-reflection about the vertex spin $\v{S}_l$ whose magnitude is 1/2. Hence, our model satisfies the condition to exclude the unique gapped phase in the recent extension of the LSM theorem to the site-reflection invariant spin chains.\cite{fuji2016, Po2017,OTT2021} Taking the continuity to the TLL phase in the limit $\lambda \rightarrow \infty$ into account, the whole nonmagnetic phase is considered to be the TLL phase. This is confirmed by the numerical diagonalization calculation of the singlet-triplet energy gap $\Delta E$. It is checked that $\Delta E$ approximately scales with the system size $L$ as $\Delta E \sim 1/L$ as shown in Fig.\ref{fig:scgap}(a) for $\delta=-0.3$ and (b) for 0.3. Similar analyses are also carried out for several other values of $\delta$. In the vicinity of the PF phase indicated by the shaded area of Fig. \ref{fig:phase}, however, the deviation from the scaling relation $\Delta E \sim 1/L$ is significant as shown in Fig. \ref{fig:scgap}(c). Nevertheless, this area shrinks with the system size. Hence, it is likely that the whole nonmagnetic phase is a TLL phase. It should be also remarked that the nonmagnetic ground state for $\delta=0$ is also a TLL phase\cite{hida2021} since it is exactly mapped onto the ground state of the spin-1/2 antiferromagnetic Heisenberg chain. For $\lambda \gtrsim \lambdac$, where $\lambdac$ is the nonmagnetic-ferrimagnetic transition point, the ferrimagnetic state with total spin $\stot >1$ comes down resulting in the level-crossing with the nonmagnetic state at $\lambdac$. Their energies measured from the ground state are plotted by the filled symbols in Fig. \ref{fig:scgap}(c). These ferrimagnetic states are macroscopically different from the nonmagnetic state in the thermodynamic limit and cannot be regarded as elementary excitations in the nonmagnetic state, even though they have the next-lowest energy for finite-size systems. Unfortunately, in the region where this type of state has the next-lowest energy, it is difficult to identify the singlet-triplet gap, since we can calculate only several lowest eigenvalues by the Lanczos method we employ in this work.

\section{Summary and Discussion}
The ground-state phases of mixed diamond chains (\ref{hama}) are investigated numerically and analytically. For comparison, the ground-state phase diagram of the corresponding classical model is calculated analytically. In the quantum case, the ground-state phase diagram is determined using the numerical exact diagonalization and DMRG method in addition to the perturbation analysis for various limiting cases. The ground state of the present model has a rich variety of phases such as the two kinds of QF phases with $\msp=1$, the QF phase with spontaneous translational symmetry breakdown, the PF phase, and the nonmagnetic TLL phase.

The fate of the infinite series of QF phases observed for $\delta=0$ is also investigated numerically. It turned out that the QF phases with large spatial periodicities $p$ are easily destroyed by small $\delta$ and replaced by the PF phase. The interpretation of this behavior is also discussed using the bosonization argument.

In the nonmagnetic phase, the unique gapped ground state is excluded based on the recently extended LSM theorem.\cite{fuji2016,Po2017,OTT2021} Combined with the numerical calculation of the energy gap, this region is considered to be the TLL phase.

So far, the experimental materials corresponding to the present mixed diamond chain are not available. However, considering the rich variety of ground-state phases, the experimental realization of the present model is expected to produce a fruitful field of quantum magnetism. With the recent progress in the synthesis of mixed spin compounds\cite{yama}, we expect the realization of related materials in the near future.
\acknowledgments

The numerical diagonalization program is based on the TITPACK ver.2 coded by H. Nishimori. Part of the numerical computation in this work has been carried out using the facilities of the Supercomputer Center, Institute for Solid State Physics, University of Tokyo, and Yukawa Institute Computer Facility at Kyoto University.

\end{document}